\documentclass[preprintnumbers,showpacs,letterpaper,aps,prd,twocolumn,superscriptaddress]{revtex4}
\usepackage{graphicx}% Include figure files
\usepackage{amsmath,amssymb,latexsym}

\usepackage{color}

\newcommand{\bfm}[1]{\mbox{\boldmath$ #1 $}} 

\begin{document}
\title{The need for a local source of UHE CR nuclei}

\author{Andrew~M.~Taylor}
\affiliation{ISDC, Chemin d'Ecogia 16, Versoix, CH-1290, SWITZERLAND\vspace{-\parskip}}

\author{Markus~Ahlers} 
\affiliation{C.N.~Yang Institute for Theoretical Physics, SUNY at Stony Brook, Stony Brook, NY 11794-3840, USA\vspace{-\parskip}}

\author{Felix~A.~Aharonian}
\affiliation{Dublin Institute for Advanced Studies, 5 Merrion Square, Dublin 2, IRELAND\vspace{-\parskip}}
\affiliation{Max-Planck-Institut f\"ur Kernphysik, Postfach 103980, D-69029 Heidelberg, GERMANY}
             
\begin{abstract}
Recent results of the Pierre Auger (Auger) fluorescence detectors indicate an increasingly heavy composition of ultra-high energy (UHE) cosmic rays (CRs). Assuming that this trend continues up to the highest energies observed by the Auger surface detectors we derive the constraints this places on the local source distribution of UHE CR nuclei. Utilizing an analytic description of UHE CR propagation we derive the expected spectra and composition for a wide range of source emission spectra. We find that sources of intermediate-to-heavy nuclei are consistent with the observed spectra and composition data above the ankle. This consistency requires the presence of nearby sources within $60$~Mpc and $80$~Mpc for silicon and iron only sources, respectively. The necessity of these local sources becomes even more compelling in the presence nano-Gauss local extragalactic magnetic fields.
\end{abstract}

\preprint{YITP-SB-11-21}

\newcounter{pub}
\newcounter{pub2}


\maketitle

\section{Introduction}\label{intro}

Ultra-high energy CRs with energies above $10^{19}$~eV arrive at Earth
with a frequency of less than
one event per square-kilometer-year
({\it i.e.}~with an energy flux of 30~eV~cm$^{-2}$~s$^{-1}$) in $\pi$ steradian. The recently completed
Pierre Auger observatory \cite{Abraham:2004dt} with a surface detector covering an area of about $3000$~km$^{2}$ is thus able to detect up to several hundred UHE CR events per year.
However, due to the steeply falling CR spectrum the arrival frequency drops by about two orders of magnitude as we go up in energy by one decade, leaving only a few events per year detectable at energies around $10^{20}$~eV. Hence, the statistical uncertainty associated with the upper end of the spectrum is still large, limiting our knowledge of UHE CR composition and origin.

Before their arrival, UHE CRs must propagate across the astronomical distance 
between their source and Earth. The relevant interactions of UHE CR nuclei during propagation are 
Bethe-Heitler pair production and photo-disintegration in collisions with photons of the cosmic background radiation. The cross-sections of both these processes rise quickly above the threshold values of about 1~MeV and 10~MeV, respectively, in the nuclei's rest-frame.
At even higher energies above 150~MeV pion production
turns on and becomes the dominant energy loss process. 
However, for the UHE CR cutoff energies and composition we consider, which are motivated by the
most recent Auger results, this process never plays a dominant role and may be safely neglected.

Provided the propagation time from their sources to Earth is greater than their 
energy loss time, UHE CRs invariably undergo these energy loss interactions. The break-up of nuclei via photo-disintegration produces lower mass nuclei with the same Lorentz factor. For heavy nuclei, the most dominant transitions are one-nucleon and two-nucleon losses. Secondary heavy nuclei remain close to the photo-disintegration resonance and quickly disintegrate further to lighter nuclei. Hence, on resonance, the initial mass composition of the sources is quickly shifted to lower atomic number values and in general shows a strong dependence on CR energy.

The arriving flux from an ensemble of UHE CR sources can be expected to contain
suppression features at the high energies at which the photo-disintegration processes turn on and the
nuclei particle's attenuation length decreases. 
Most prominently, for the case of a proton-dominated spectrum, the flux is expected to be suppressed 
by the {\it Greisen-Zatsepin-Kuz'min} (GZK) cutoff~\cite{Greisen:1966jv,Zatsepin:1966jv} due to resonant 
pion photo-production interactions with the cosmic microwave background (CMB).
Intriguingly, a suppression of the CR spectrum at about $5\times10^{19}$~eV has been observed at a statistically significant level~\cite{Abbasi:2007sv,Abraham:2008ru}.
However, since a similar feature may also appear in the spectrum from nuclei primaries, the
observation of such a feature provides little clue as to the underlying composition.
Such a suppression feature becomes a cutoff in the arriving flux if the attenuation length drops below the distance to the nearest UHE CR source. Thus, the shape of the suppression/cutoff feature does contain 
information about the source distribution, as has been investigated already for the 
case of UHE CR protons~\cite{Aharonian:1994nn,Berezinsky:2002nc,Taylor:2008jz}.

On their arrival at Earth, UHE CRs interact with molecules in the atmosphere and deposit their energy in the form of extensive air showers. The characteristics of these showers along the shower depth $X$ (in g/cm${^2}$) contain vital UHE CR composition information.
On average, proton-induced showers reach their maximum development, 
$\langle X_{\rm max} \rangle$, deeper in the
atmosphere than do showers of the 
same energy generated by heavier nuclei. Accompanying this effect, the shower to 
shower fluctuation of $X_{\rm max}$ about the mean, $\mathrm{RMS}(X_{\rm max})$, is 
larger for proton-induced showers than for iron-induced showers of the same energy. 
As a result, measurements of both $\langle X_{\rm max} \rangle$ and 
$\mathrm{RMS}(X_{\rm max})$ can be used to infer the average chemical composition of 
the UHE CRs as a function of energy.

Auger has now released their first measurement results of 
$\mathrm{RMS}(X_{\rm max})$~\cite{Unger:2011ry,Abraham:2010yv}, along with those of 
$\langle X_{\rm max} \rangle$. 
These results seem to imply that the UHE CR spectrum 
contains a large fraction of intermediate or heavy mass nuclei, becoming 
increasingly heavy at high energies ($\sim 10^{19.5}$~eV). 
Furthermore, the small values of $\mathrm{RMS}(X_{\rm max})$ measured 
by Auger also imply that UHE CRs are composed of species with a relatively narrow 
distribution of charge at the highest measured energies, containing little or no protons 
or light nuclei. 
In this way, the new $\mathrm{RMS}(X_{\rm max})$ measurements not only 
confirm and reinforce the conclusions drawn from their earlier average depth of shower 
maximum measurements, but also provide complementary information that enables one 
to constrain the distribution of the various chemical species present within the 
UHE CR spectrum.
Curiously, recent corresponding measurements 
by the Telescope Array UHE CR detector do not appear to agree with the Auger results
\cite{Matthews:2011zz}, and instead seem to be consistent with a light composition 
as was indicated by its predecessor HiRes~\cite{Abbasi:2009nf}. 
Though an understanding of this conflict is crucial for future progress in the
field, we here chose to adopt only the Auger results.

Interestingly, the intermediate-heavy nuclei scenario emerging from these 
measurements may find consistency with the presently observed mild departure
from isotropy and absence of point sources observed at UHE \cite{:2010zzj}. 
This could occur through the introduction of $\sim 10^{\circ}$ deflections in the 
turbulent ($\mu$G) Galactic magnetic field structure \cite{Giacinti:2011uj} or in 0.1~nG 
turbulent extragalactic field structure \cite{Dolag:2004kp}, from an anisotropic 
distribution of nearby sources. Such large deflections would limit the 
prospects for future UHE CR astronomy for all but the brightest and closest sources.
The correlation of UHE CR with nearby objects, however, still awaits to be resolved 
within an intermediate-heavy nuclei framework, and deserves further consideration.

Following our previous investigations into the composition of UHE CRs~\cite{Hooper:2009fd}, 
an intermediate-heavy (silicon-to-iron) type composition was found to be 
motivated by the present complete Auger data set, with a hard ($\alpha< 2$) 
injection spectral index \footnote{This spectral index refers to an injection spectrum of the form ${\rm d}N/{\rm d}E\propto E^{-\alpha}$.}. 
\setcounter{pub}{\value{footnote}}
Furthermore, this agreement received only very mild improvements by the
additional consideration of an admixture of species. Building further on these 
results, we here focus on the local UHE CR source distribution inferred
to exist if the observed trend in the composition continues up to highest 
energies observed by the ground array ($\sim 10^{20.2}$~eV). 
By considering simple silicon and iron source composition scenarios, we
investigate how far away these sources can afford to be without detrimentally
damaging the agreement with the Auger data.
To aid this
investigation we utilise an analytic description of UHE CR nuclei propagation
already developed by the authors~\cite{Ahlers:2010ty}.

\section{UHE CR Nuclei Fluxes from a Uniform Distribution of Sources}\label{Uniform_Dist}

With no prior knowledge about the UHE CR source population, a ``universal'' 
(homogeneous and isotropic) distribution is generally assumed. We adopt this 
assumption here as a means of investigating signatures of a departure from it.
In order to quantify the effect of a different source distribution in this paper, we 
separate out the fluxes produced from source regions with shells of radii 0-3~Mpc, 
3-9~Mpc, 9-27~Mpc, 27-81~Mpc, and 81-243~Mpc surrounding the Earth.
In this way, the 
results obtained may be used to encapsulate the effects introduced by a 
non-``universal'' local void of UHE CR sources.

\begin{figure}[h!]
\begin{center}
{\includegraphics[angle=0,width=0.9\linewidth,type=pdf,ext=.pdf,read=.pdf]{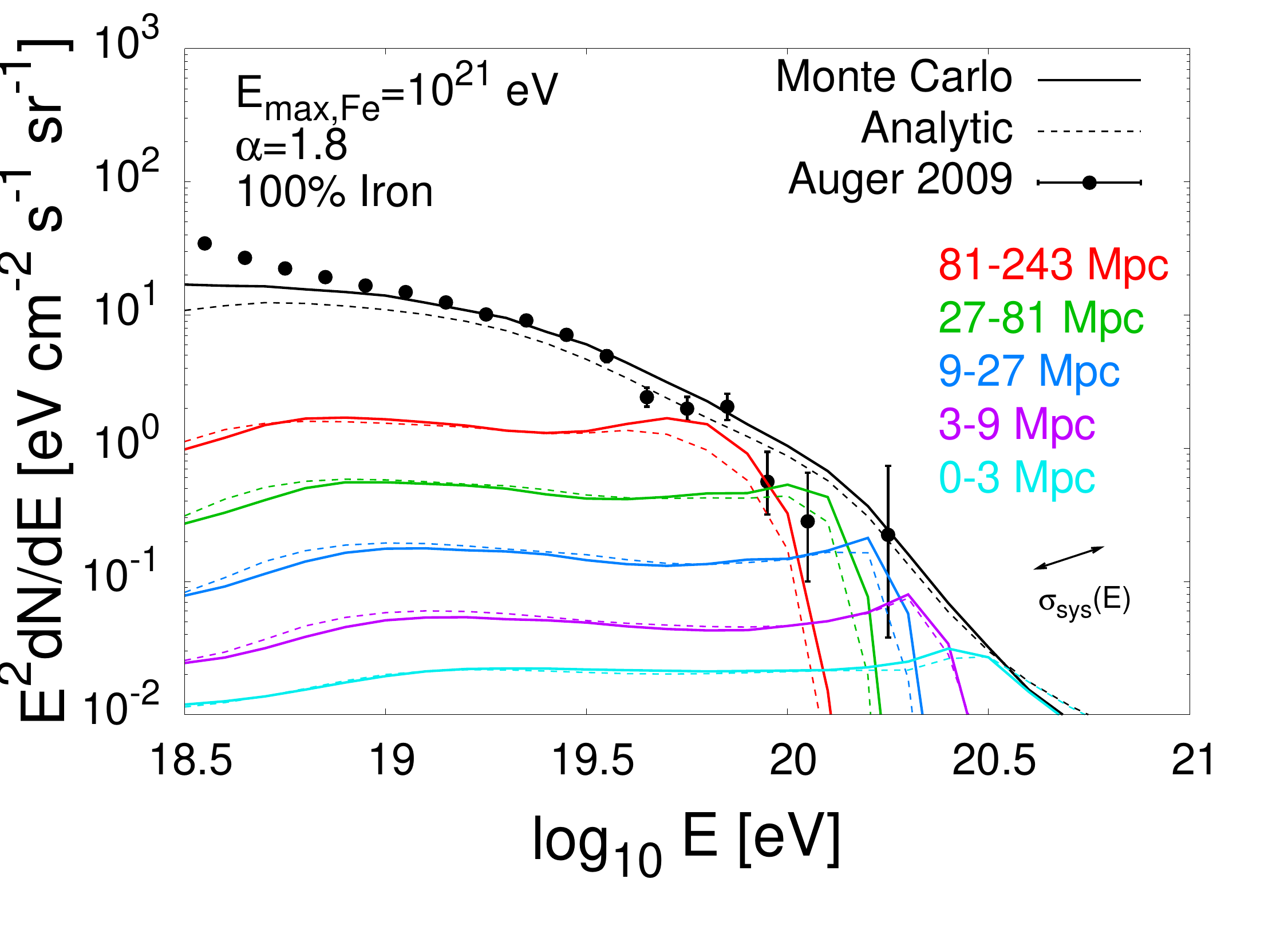}}
{\includegraphics[angle=0,width=0.9\linewidth,type=pdf,ext=.pdf,read=.pdf]{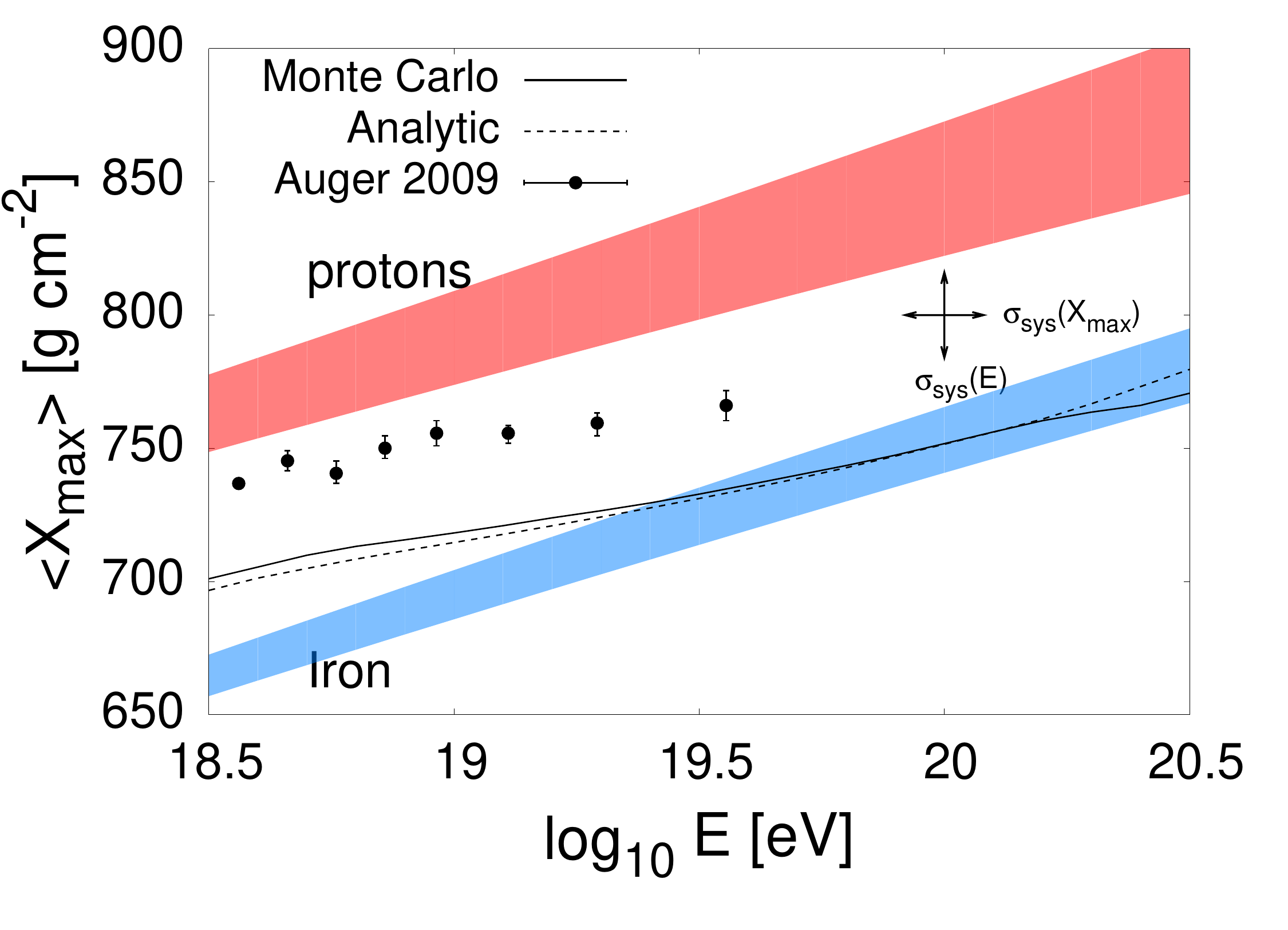}}
{\includegraphics[angle=0,width=0.9\linewidth,type=pdf,ext=.pdf,read=.pdf]{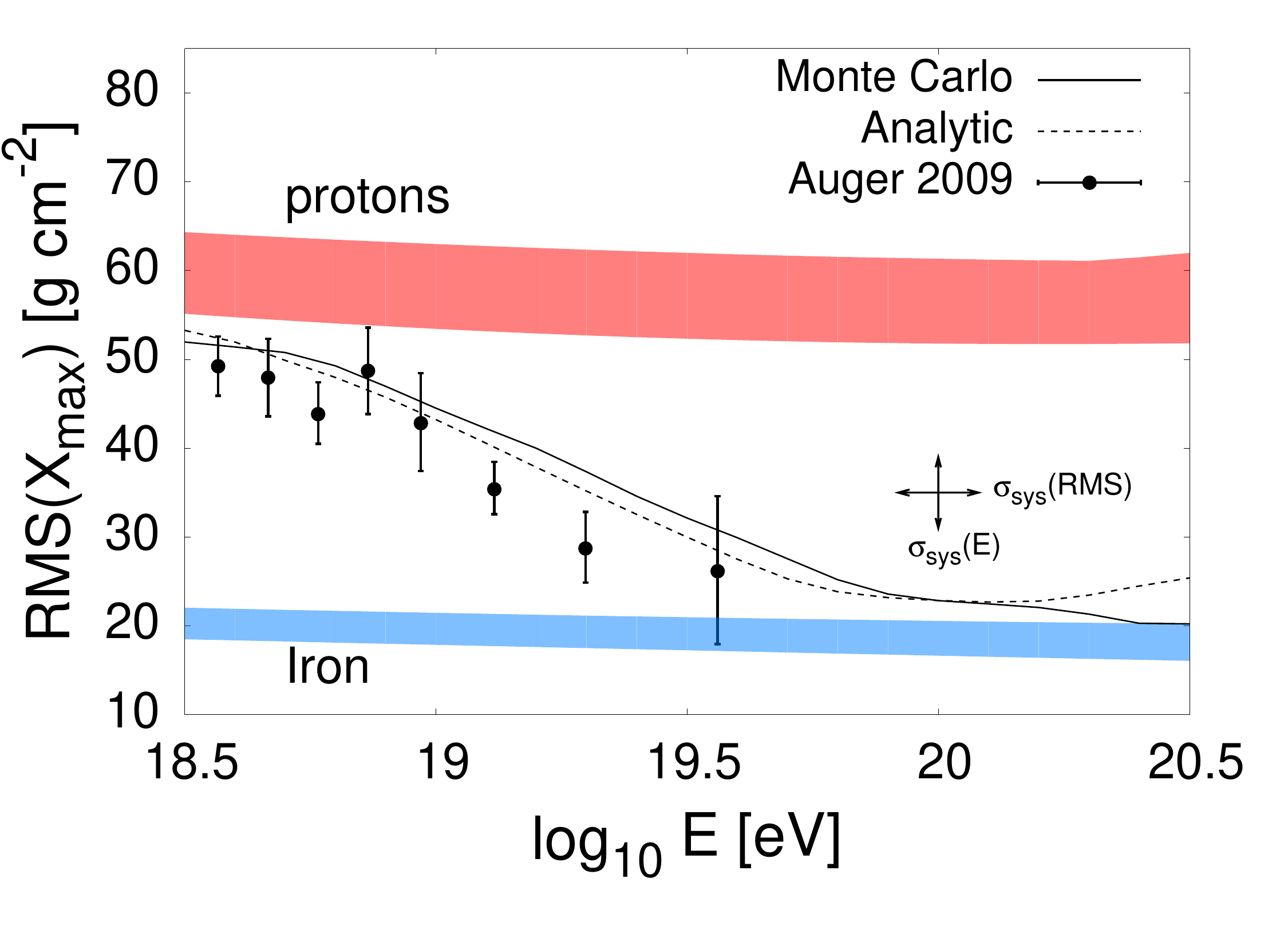}}
\caption[]{A comparison of the Monte Carlo and analytic flux results following
the emission of Fe nuclei from sources with $E_{\rm Fe, max}=10^{21}$~eV and $\alpha=1.8$, to recent Auger measurements \cite{Abraham:2009wk,Unger:2011ry,Abraham:2010yv}.
{\bf Top panel:} A breakdown of the arriving flux from shells of UHE CR Fe sources. {\bf Middle and bottom panels:} The arriving flux's corresponding $\langle X_{\rm max} \rangle$ and $\mathrm{RMS}(X_{\rm max})$ values, which describe the composition. The hadronic model QGSJET~11~\cite{qgsjet_11} has been adopted for these comparative calculations. The red  and blue bands show the range of predictions of the $\langle X_{\rm max} \rangle$ and $\mathrm{RMS}(X_{\rm max})$ values for protons and iron, for various hadronic interaction models \cite{qgsjet_11,qgsjet,sibyll,epos}. In all three panels the arrows labeled $\sigma_{\rm sys}$ depict the size of the Auger systematic errors.}
\label{breakdown_analytic}
\end{center}
\end{figure}

We show in the upper-panel of Fig.~\ref{breakdown_analytic} a breakdown of the total 
arriving flux from an ensemble of sources assuming they have a homogeneous distribution
and emit a purely iron type UHE CR nuclei composition.
These results have been obtained using both a Monte Carlo description of UHE CR nuclei 
propagation, whose details are described in Ref.~\cite{Hooper:2006tn}, as well as an 
analytic description whose details are described below. An injection spectrum of 
$\alpha=1.8$ and a maximum energy of $E_{\rm Fe, max}=10^{21}$~eV have here been 
assumed \footnote{This maximum energy refers to the charge-dependent exponential cutoff energy, 
${\rm d}N/{\rm d}E\propto E^{-\alpha}e^{-E/E_{Z,\rm max}}$, where $E_{Z,\rm max}=(Z/26)\times E_{\rm Fe, max}$.},
\setcounter{pub2}{\value{footnote}}
We use the iron equivalent value of the cutoff, $E_{\rm Fe, max}$, as the reference value
since this allows a rigidity independent comparison of the cutoffs
values for the type species we consider.
The injection spectrum values adopted in Fig.~\ref{breakdown_analytic} were motivated by 
previous investigations into UHE CR nuclei sources~\cite{Hooper:2009fd}. From this figure, 
it is clearly seen that the arriving flux at the highest UHE CR energies ($E>10^{20}$~eV) 
is dominated by the local source population.

In the middle and bottom panels of Fig.~\ref{breakdown_analytic} the composition
related information of the arriving flux,  $\langle X_{\rm max} \rangle$ and 
$\mathrm{RMS}(X_{\rm max})$, are shown. For these results, the hadronic interaction model 
QGSJET~11~\cite{qgsjet_11} has been adopted. The broad red and blue lines depicting the
``proton'' and ``iron'' values in these plots indicate the spread of predicted values 
from a range of hadronic interaction models~\cite{qgsjet_11,qgsjet,sibyll,epos}.
The departure of the black line from the blue region in both these panels thus
demonstrates the degree of photo-disintegration incurred by the injected iron nuclei 
en route in the extragalactic radiation fields.

To simplify the picture the effects 
of the presence of extragalactic magnetic fields on UHE CR propagation are neglected in 
this section. In section~\ref{Magnetic_Fields}, these effects are incorporated to see how 
the field-free results in this section are altered.

The shape of the cutoff feature in the UHE CR spectrum carries within it valuable 
information about the local source distribution. A simple analytic description 
of this feature, found in Ref.~\cite{Hooper:2008pm}, and developed further in 
\cite{Ahlers:2010ty}, allows the different cutoffs 
in the fluxes from the various shells to be easily interpreted. To first order, photo-disintegration of heavy nuclei can be approximated by one-nucleon loss. The flux of nuclei with mass number $A$ from a source at distance $L$ with initial mass number $A_{\rm ini}$ follows simply,
\begin{eqnarray}
\frac{N_{A} (E_{A},L)}{N_{A_{\rm ini}} (E,0)} = \sum_{m=A}^{A_{\rm ini}} l_{0} l_m^{A_{\rm ini}-1} 
e^{-{\frac{L}{l_m}}}\!\!\prod_{p=0(\neq m)}^{A_{\rm ini}} \frac{1}{l_m -l_p}\,.
\label{distribution} 
\end{eqnarray}
Here, $l_a$ denotes the interaction length of one-nucleon loss of the nucleus with mass number $a$.
By employing (\ref{distribution}), and summing over all 
possible mass numbers $A$ (from $1$ to $A_{\rm ini}$) different nucleon loss contribution 
functions for the flux from the source, the total nuclei flux from a single source is obtained,
\begin{eqnarray}
\frac{{\rm d}N_{\rm total} (E,L)}{{\rm d}L}=\sum_{A=1}^{A_{\rm ini}}\frac{N_{A} (E_{A},L)}{N_{A_{\rm ini}} (E_{A_{\rm ini}},0)}.
\end{eqnarray}

The agreement between the Monte Carlo and analytic results, seen in Fig.~\ref{breakdown_analytic},
demonstrates that for negligible ($<$pG) extragalactic fields the local source 
information (spectrum and composition) is essentially encapsulated by our simplified analytic model. 
Indeed, the different sets of photo-disintegration cross-sections used for the Monte Carlo and 
analytic descriptions, for which \cite{Khan:2004nd} and \cite{TALYS} have been used respectively, 
demonstrate further the robustness of these results. Using this 
analytic description, sets of models may be quickly scanned through in order to find 
the optimum set of injection index $\alpha$, exponential energy cutoff $E_{\rm max}$ and source composition 
values able to ``best-fit'' the Auger results above $10^{19}$~eV.

\begin{figure}[ht]
{\includegraphics[angle=0,width=0.9\linewidth,type=pdf,ext=.pdf,read=.pdf]{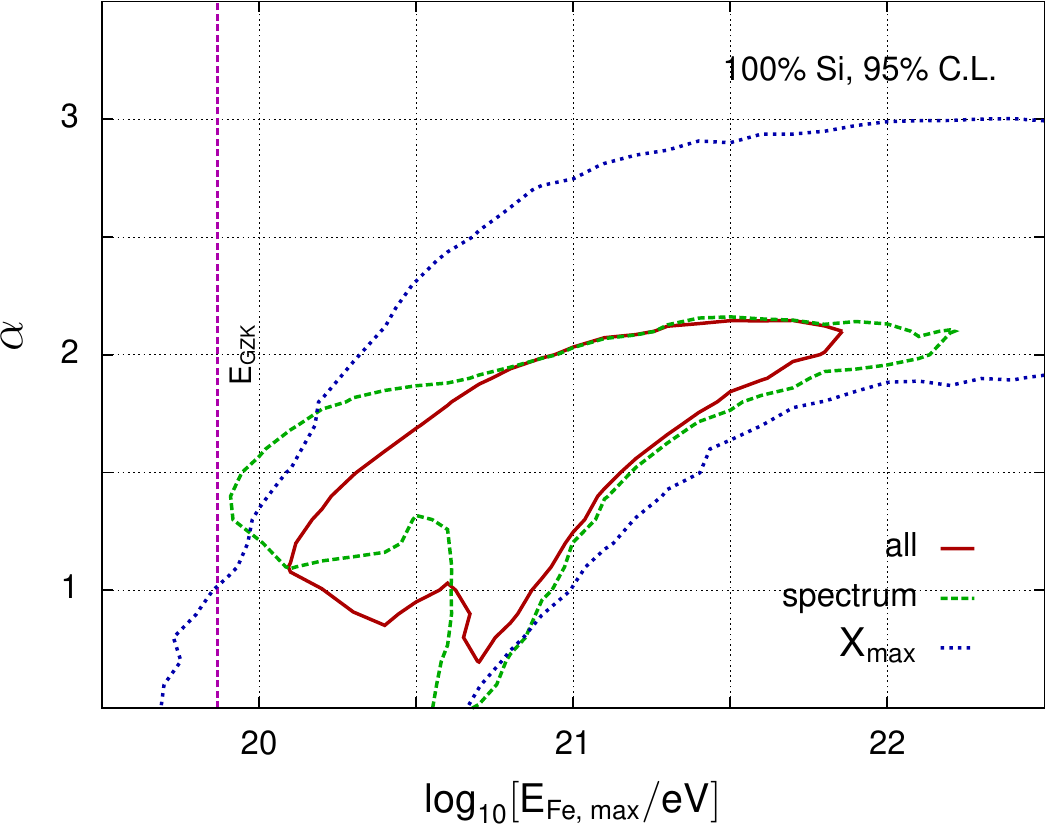}}\\
\vspace{5mm}
{\includegraphics[angle=0,width=0.9\linewidth,type=pdf,ext=.pdf,read=.pdf]{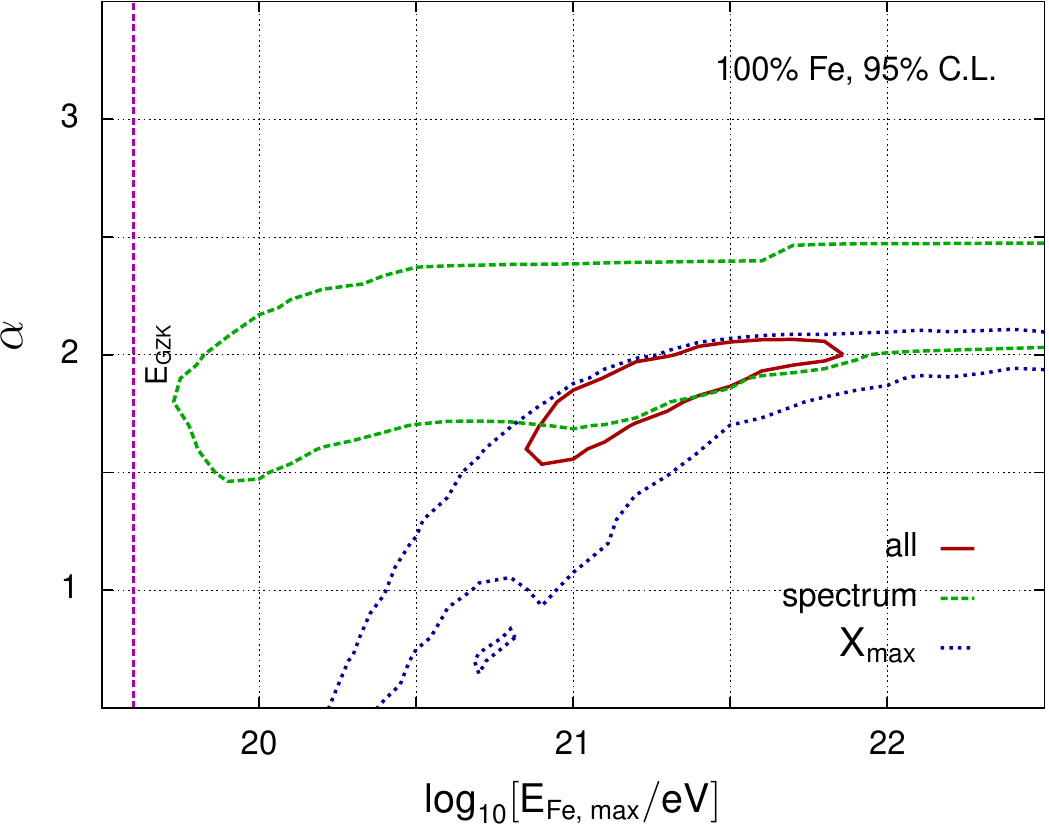}}
\caption{Contour plots showing both the spectrum and $X_{\rm max}$ (ie. both $\langle X_{\rm max} \rangle$ and 
$\mathrm{RMS}(X_{\rm max})$) goodness-of-fit results for silicon (upper panel) and iron (lower panel) source compositions. The spectral fits to the data have been carried out for energies above $10^{19}$~eV, and the QGSJET~11 hadronic model has been used as the hadronic model needed to obtain the theoretical $\langle X_{\rm max} \rangle$ and $\mathrm{RMS}(X_{\rm max})$ values.}
\label{Contours}
\end{figure}

Through the application of a goodness-of-fit (GOF) test, taking into account the systematic uncertainties 
in the measurement of CR energy, $\langle X_{\rm max} \rangle$, and $\mathrm{RMS}(X_{\rm max})$  
(see Appendix~\ref{gof_test}), whose sizes are shown in Fig.~\ref{breakdown_analytic}, we find that, in agreement with our earlier investigations~\cite{Hooper:2009fd}, 
a intermediate-heavy composition ($A>20$), hard spectral indices 
($\alpha<2$) and intermediate type cutoff energies 
($E_{\rm Fe, max}\sim 10^{21}$~eV) are best able to describe the current data. 
Contour plots showing the ``best-fit'' regions for both silicon and iron only type
sources are shown in Fig.~\ref{Contours}. 
For lighter nuclei, the contour space was found to be considerably diminished, with no 
such contours existing for the proton-only scenario, even at the 99\% C.L.
We also indicate in the plots the position
of the GZK energy $E_{\rm GZK}$ in terms of $E_{\rm Fe,max}$. The fit to the data prefers 
considerably higher cutoff energies than the GZK energy. 
These plots demonstrate that both the spectra and composition information both provide 
new and differing constraints on the source spectral parameters, particularly for the case 
of heavy nuclei (iron) type sources.

\section{The Effect of the Finite Distance to the Nearest Source}
\label{First_Source}

The number density of local sources can not be much smaller than $10^{-5}$ Mpc$^{-3}$ as can be estimated, {\it e.g.}, from the absence of ``repeaters'' in CR data~\cite{Blasi:2003vx,Kashti:2008bw}. It is hence expected that the flux of CRs from sources much larger than a few 100~Mpc can be well approximated by a homogeneous distribution of sources.
However, the finite (non-zero) distance to the nearest source of UHE CR nuclei leads to a breakdown of
the homogeneous source distribution spectrum result at the highest energies, as already indicated
in Fig.~\ref{breakdown_analytic}.
We here demonstrate the effect introduced by this local void of UHE CR sources using both
concrete examples as well as a statistical approach. 

\begin{figure}[ht]
\begin{center}
{\includegraphics[angle=0,width=0.9\linewidth,type=pdf,ext=.pdf,read=.pdf]{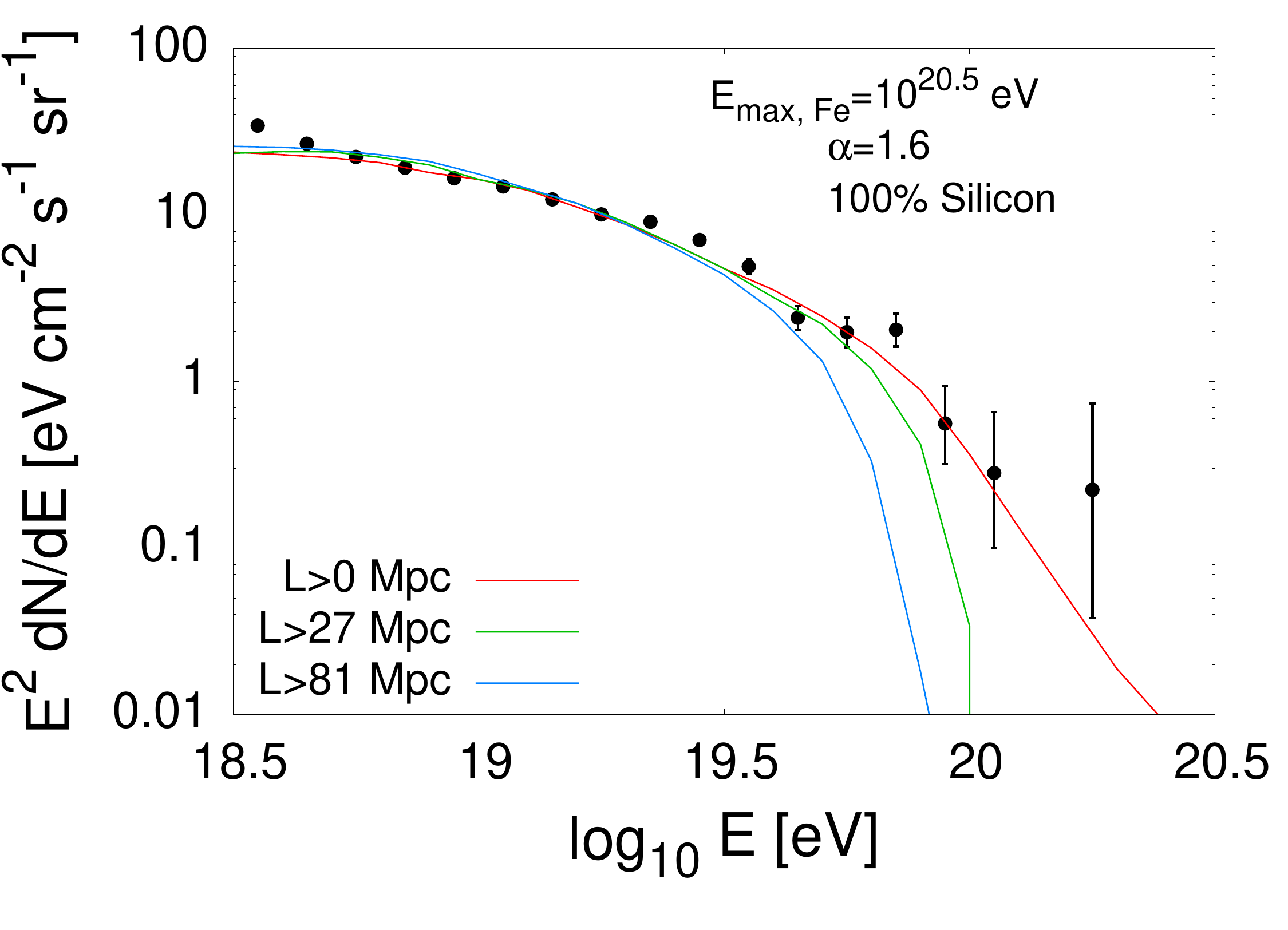}}\\
{\includegraphics[angle=0,width=0.9\linewidth,type=pdf,ext=.pdf,read=.pdf]{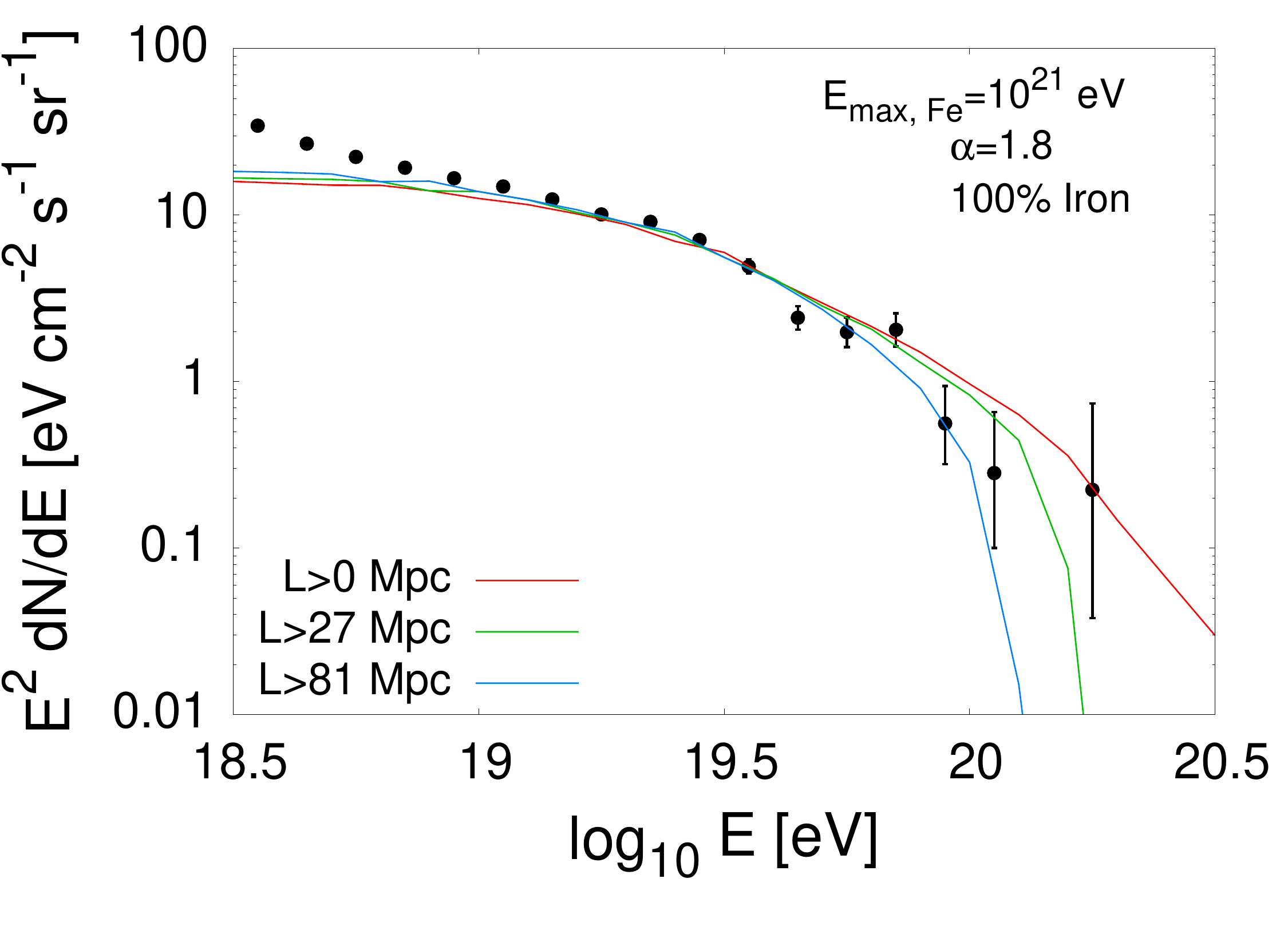}}\\
\caption{The arriving UHE CR flux from a uniform distribution of UHE CR sources emitting a pure silicon (top), and pure iron (bottom) composition. A source spectral index ($\alpha$) and cutoff energy ($E_{\rm Fe, max}$) of $1.6$ and $10^{20.5}$~eV respectively for silicon and $1.8$ and $10^{21}$~eV respectively for iron have been adopted, as motivated by our goodness-of-fit results in the previous section.}
\label{source_dist}
\end{center}
\end{figure}

As an illustrative example, using typical ``best-fit'' model parameters motivated from our 
GOF results in the previous section, we show explicitly in Fig.~\ref{source_dist} 
the alteration to the flux due to the failure of the homogeneous approximation. 
The lower (upper) panel in Fig.~\ref{source_dist} show how the arriving flux from
an ensemble of sources emitting iron (silicon) type UHE CR is altered due to a non-zero
distance to the nearest source.
These results demonstrate the introduction of a very strong cutoff feature due to a
non-zero distance to the first source, with this feature occurring at lower energy for 
intermediate nuclei such as silicon, than for heavy nuclei such as iron.
It should be emphasised that the position (in energy) of the cutoff introduced by
the distance to the nearest source is roughly independent of the 
spectral index ($\alpha$) and cutoff energy ($E_{\rm Fe, max}$) of the
primary spectrum. However, this result is only true when the source's maximum energy sits at 
energies much larger than the cutoff energy introduced by the local void of sources.
We re-iterate here that these results rest on the assumption that the composition above 
energies where it has been measured ($10^{19.5}$~eV), continues to consist of intermediate-heavy 
nuclei up to the highest observed energy flux measurements ($10^{20.2 }$~eV).

\begin{figure}[ht]
\begin{center}
{\includegraphics[angle=0,width=0.9\linewidth,type=pdf,ext=.pdf,read=.pdf]{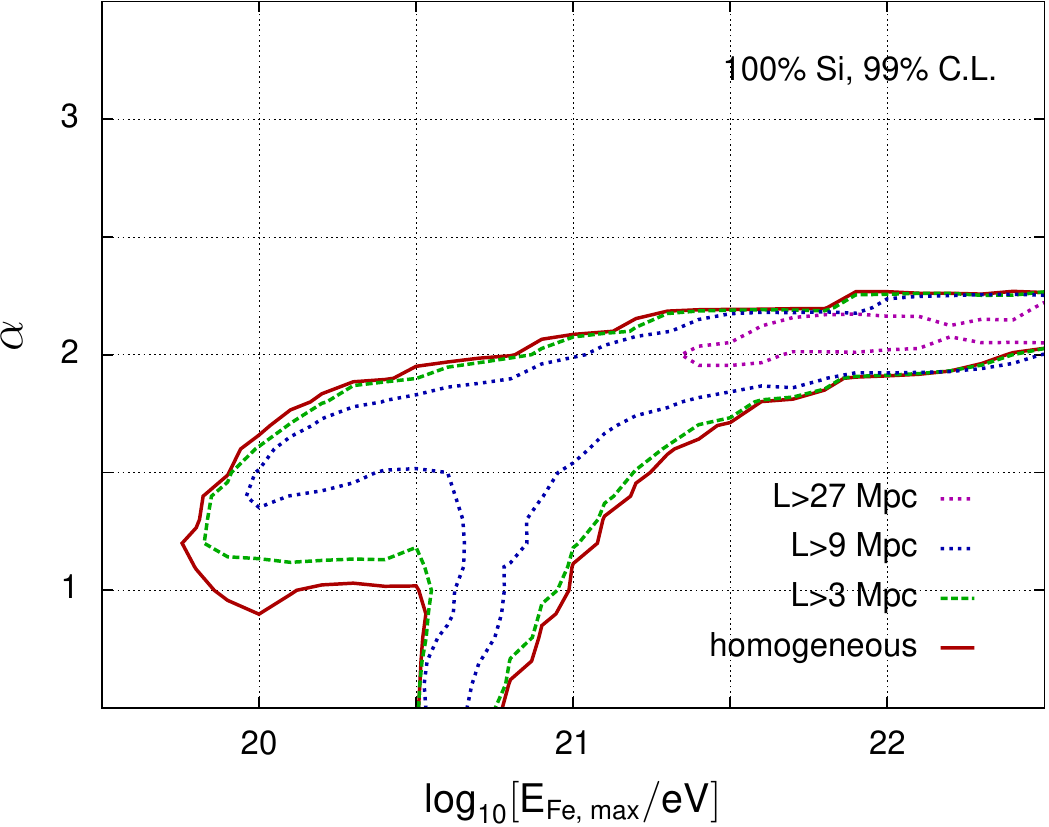}}\\
\vspace{5mm}
{\includegraphics[angle=0,width=0.9\linewidth,type=pdf,ext=.pdf,read=.pdf]{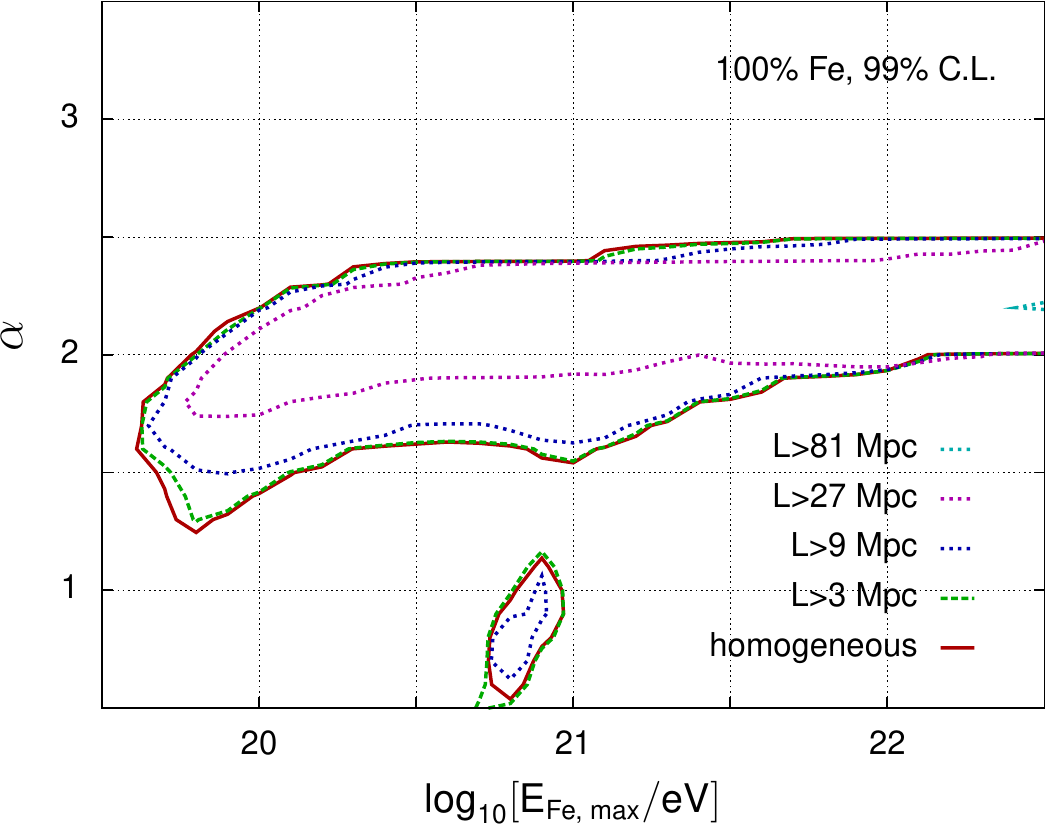}}\\
\caption{Plots showing the disappearance of the 99\% GOF contours as the distance to the first source is increased up to $81$~Mpc.}
\label{source_dist2}
\end{center}
\end{figure}

To put these results on a more general footing, we show in 
Fig.~\ref{source_dist2} the alteration to the GOF 
contour plots as the distance to the nearest source in increased.  
We find that for both silicon-only and iron-only source scenarios, the 99\% C.L. 
undergoes a rapid decrease in size for minimum source distances in the
range $9-27$~Mpc and $27-81$~Mpc respectively.
Furthermore, for larger local voids, the remaining $E_{\rm Fe, max}$ values are
larger than those typically considered feasible for candidate UHE CR sources.
By imposing the constraint that the cutoff energy, $E_{\rm Fe, max}$, sits below
$10^{22}$~eV, the subsequent upper limit on the source distance of 60~Mpc
and 80~Mpc are found for silicon and iron type source respectively, at
the 99\% C.L. This reaffirms the result shown in Fig.~\ref{source_dist}, that a 
large $\gtrsim$81~Mpc local void in the source distribution would lead to 
difficulties in finding self-consistent fits to the spectral, 
$\langle X_{\rm max} \rangle$, and $\mathrm{RMS}(X_{\rm max})$ data of Auger.

We have assumed throughout this work so far that the 
extragalactic magnetic field strength is negligible. 
Though the overall flux from the {\it total}
ensemble of shells does not vary when an isotropic and homogeneous distribution
of magnetic fields is introduced, the components arriving from different source shells
will change. Furthermore, with an aim to investigate the possibility of a
lack of nearby UHE CR sources, it is necessary that we take into account
extragalactic magnetic field effects. In the following section we investigate
these effects with the aim of making our conclusions more general.

% MENTION + JUSTIFY NEGLECTING OF EXTRA-GALACTIC MAGNETIC FIELDS

\section{The Effect of Extragalactic Magnetic Fields}
\label{Magnetic_Fields}

In this section we investigate the effect these fields 
have on the arriving UHE CR flux and composition. In our calculations,
the extragalactic magnetic field is always assumed to have a coherence length 
of 1~Mpc. Thus, within each of the magnetic patches (cells), the field contains a 
uniform component whose direction is assumed to be orientated independently (randomly) to 
that of its corresponding orientation in the neighbouring cells. 
For particles with Larmor radii smaller than the magnetic patch coherence size
({\it e.g.}~for iron nuclei with energies $\lesssim 10^{19}$~eV in a nG field),
we assume a power law distribution of magnetic turbulence of the form 
$P(k)\propto k^{-q}$ (where $k=2\pi/\lambda$), for which $q=5/3$ corresponds to a 
Kolmogorov-type spectrum, $q=3/2$ corresponds to a Kraichnan-type spectrum.
This angular (and eventually spatial) diffusion of the particles is treated following 
a method very similar to that described in \cite{Aharonian:1994nn}, which we refer
to as the ``delta-approximation'' method (described in further detail in 
Appendix~\ref{turb_fields}). A Kolmogorov-type description of
the magnetic field turbulence spectrum is assumed here.
Ultra-high energy CRs in the simulation were considered to have arrived once 
they reached a distance of 100~kpc from Earth (this length scale
being chosen to be smaller than both the corresponding particle loss length
and gyro-radius).

\begin{figure}[ht]
\begin{center}
{\includegraphics[angle=0,width=0.9\linewidth,type=pdf,ext=.pdf,read=.pdf]{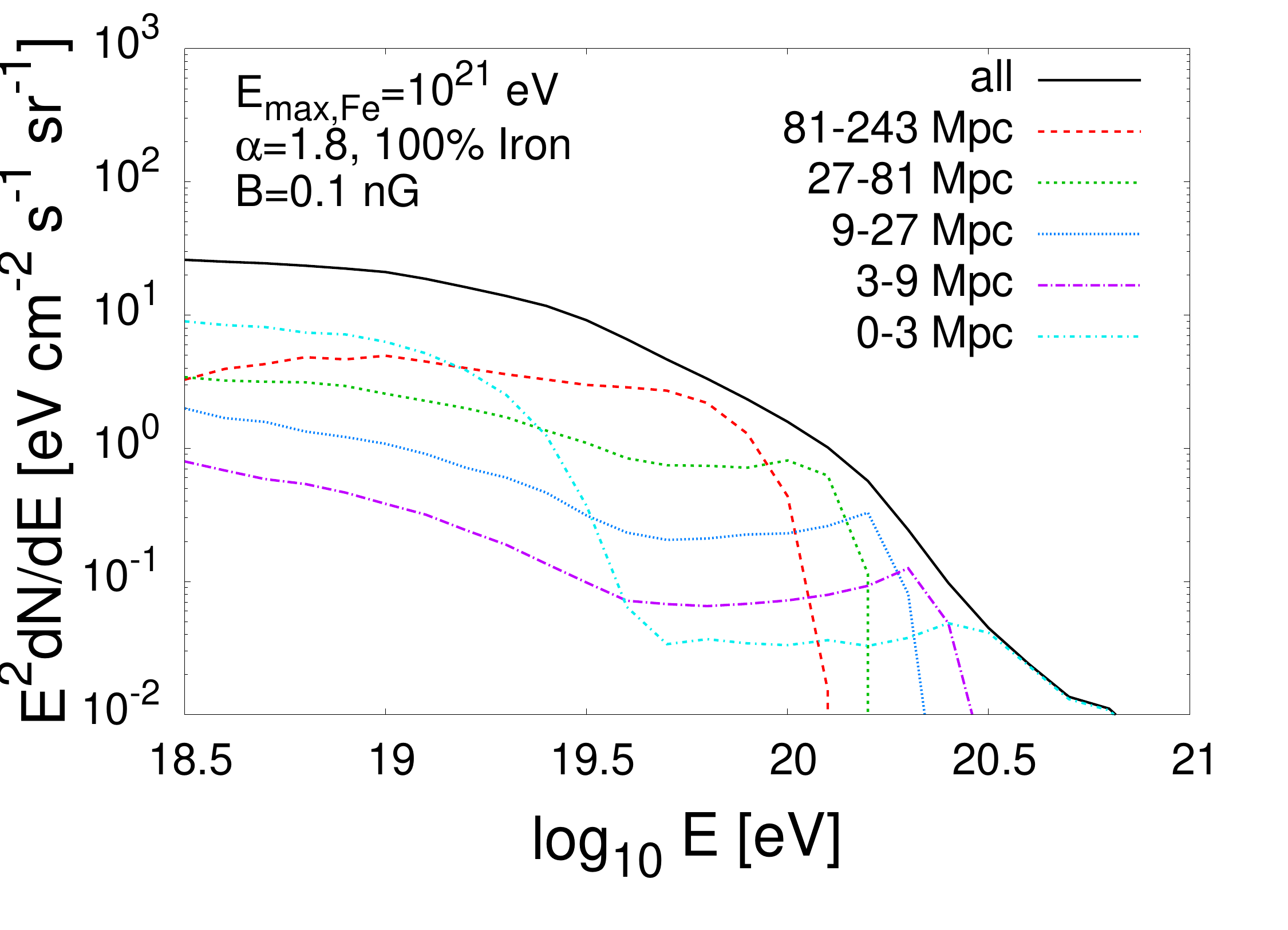}}\\
{\includegraphics[angle=0,width=0.9\linewidth,type=pdf,ext=.pdf,read=.pdf]{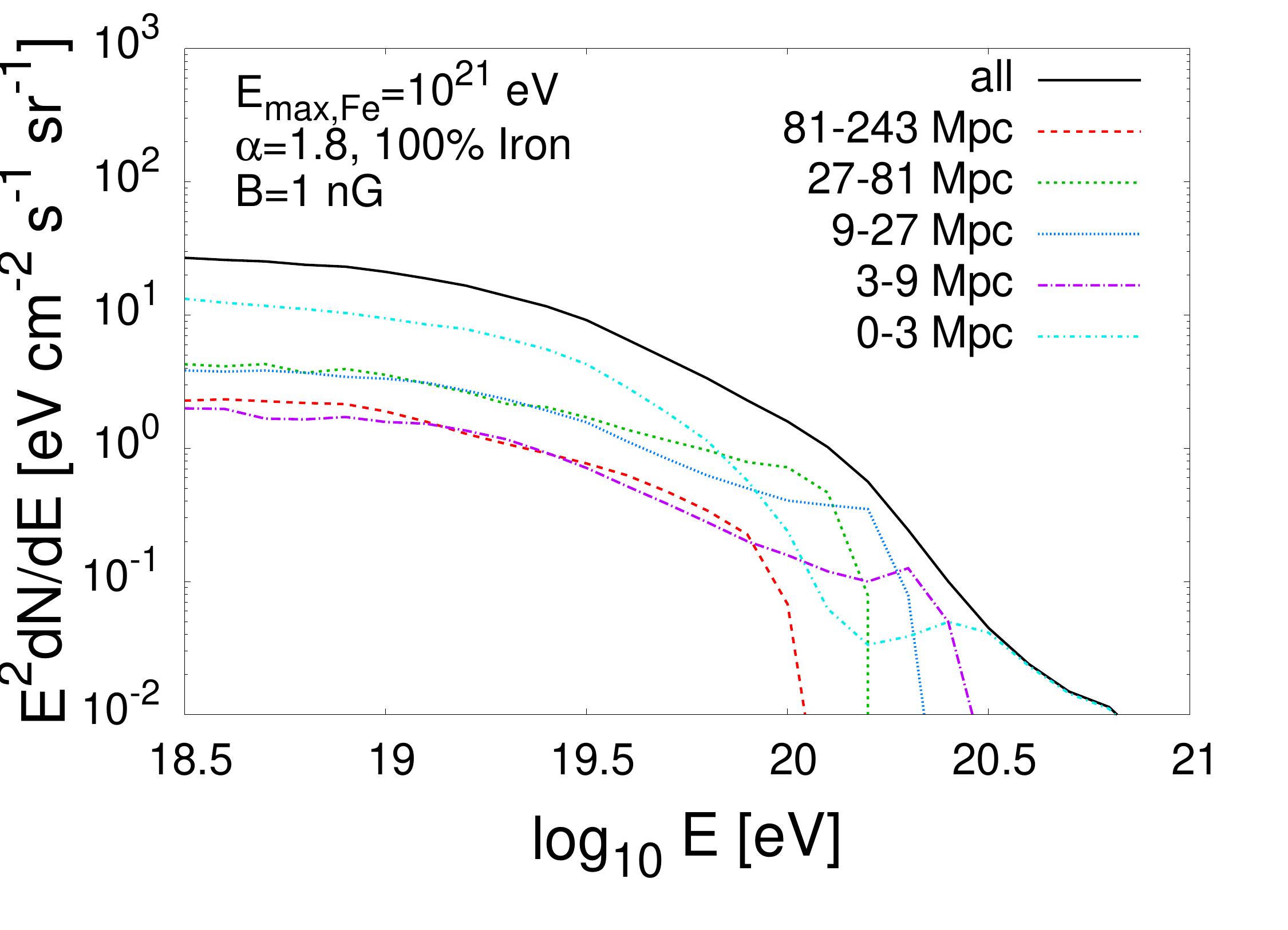}}
\caption{A breakdown of the arriving flux from shells of UHE CR Fe emitting sources with an extragalactic magnetic field of strength (top) 0.1~nG and (bottom) 1~nG. Coherence lengths of 1~Mpc have been assumed for these calculations, which were obtained using a Monte Carlo description for UHE CR nuclei propagation.}
\label{Si_Fe_shells_Mag}
\end{center}
\end{figure}

Provided that a continuous distribution of sources exists on all scales, and that the
magnetic fields are homogeneous and isotropic, extragalactic magnetic fields have no 
overall effect on the arriving flux \cite{Aloisio:2004jda}. However, since a minimum 
source distance scale must exist, a ``magnetic horizon'' is expected, with the flux 
from the nearby sources being prevented from arriving to us below a given energy
\cite{Lemoine:2004uw}. Furthermore, the contribution of sub-``magnetic horizon''
source shells can be increased by the presence of extragalactic magnetic fields,
altering somewhat the flux arriving from the different source shells.

As seen in Fig.~\ref{Si_Fe_shells_Mag}, the presence of extragalactic magnetic 
fields enhances the role played by local regions of sources on the arriving UHE CR 
flux. As envisaged from these results, and demonstrated in \cite{Globus:2007bi}, 
a low energy cutoff of the arriving flux from distant shells of sources is introduced
by the combined effect of a lack of local sources and the presence of non-negligible
extragalactic magnetic fields. Below this cutoff energy, the UHE CR flux from even 
the nearest by sources is suppressed.

\begin{figure}[ht]
\begin{center}
{\includegraphics[angle=0,width=0.9\linewidth,type=pdf,ext=.pdf,read=.pdf]{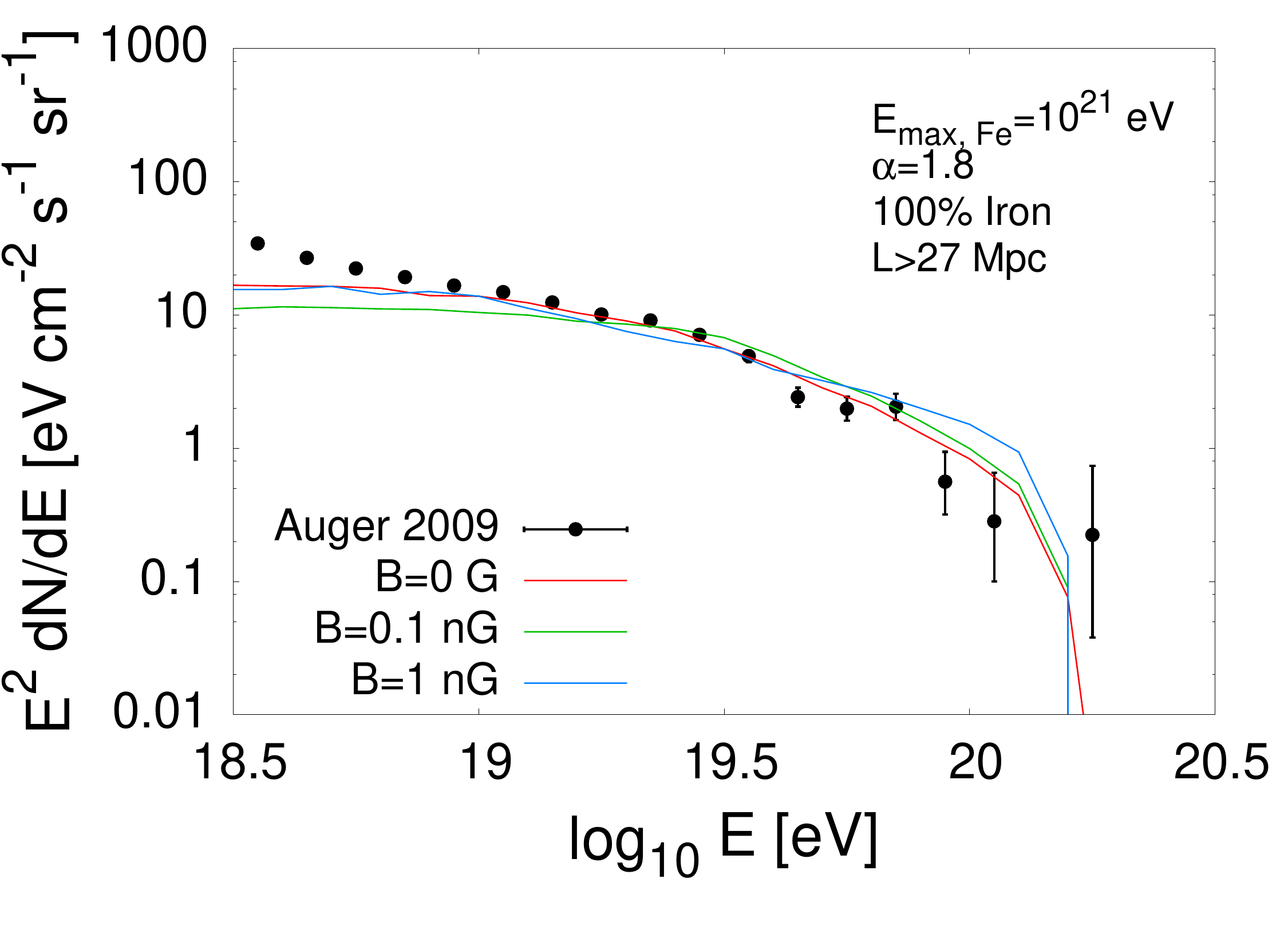}}\\
{\includegraphics[angle=0,width=0.9\linewidth,type=pdf,ext=.pdf,read=.pdf]{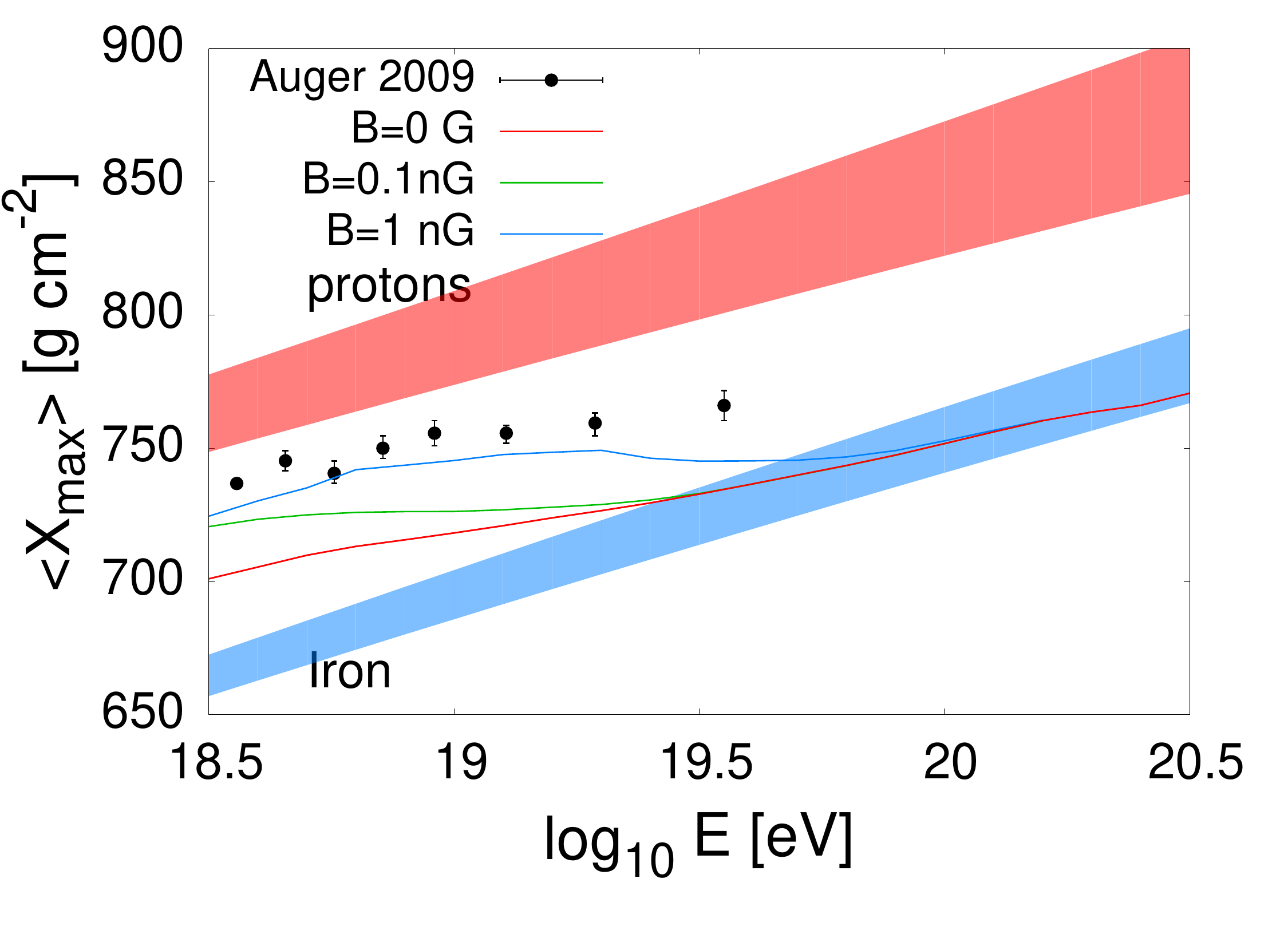}}\\
{\includegraphics[angle=0,width=0.9\linewidth,type=pdf,ext=.pdf,read=.pdf]{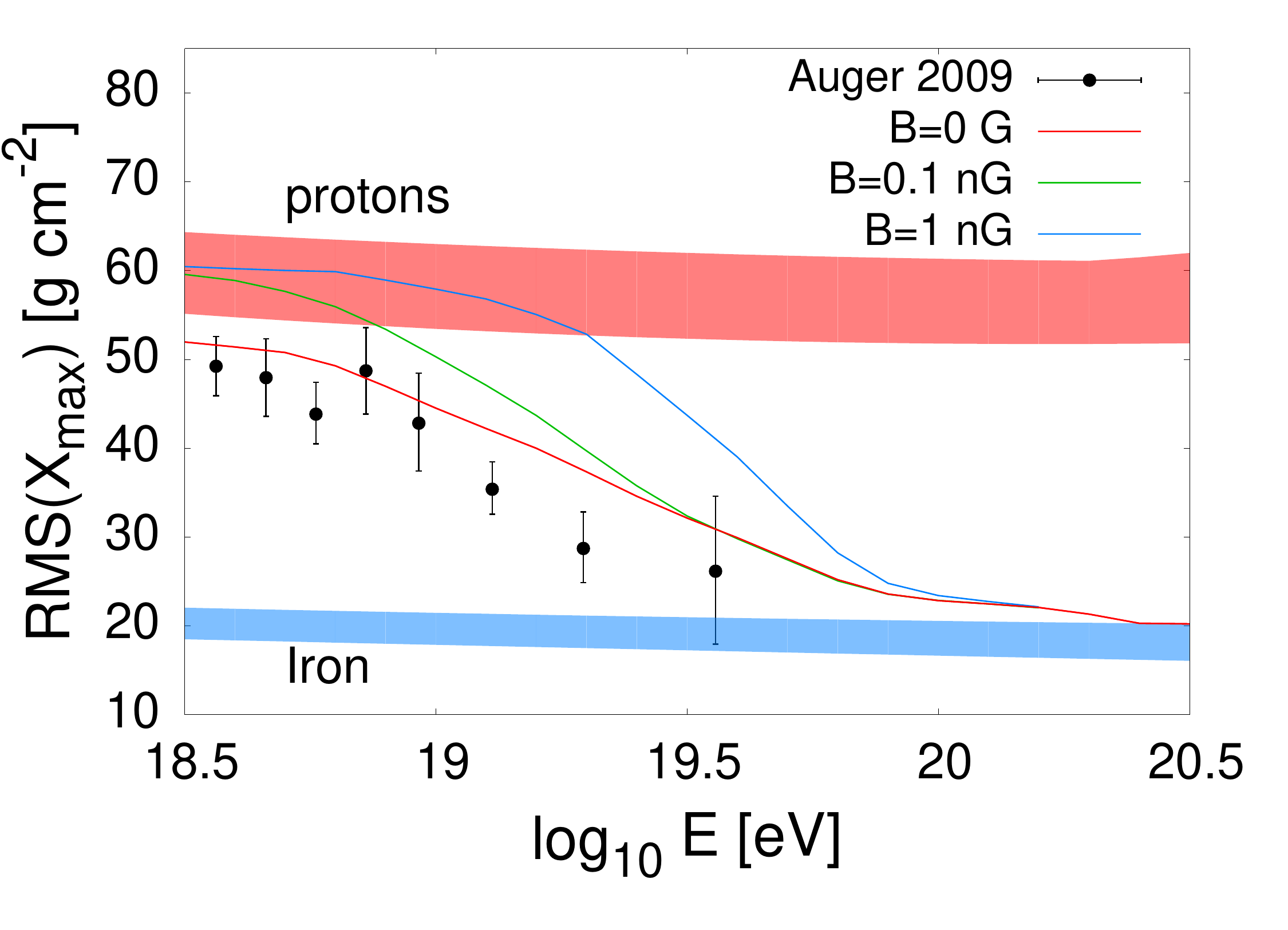}}
\caption{The combined effect on the iron spectrum (top panel) and composition (middle and bottom panel) due to a local void of sources (L$>27$~Mpc) and a non-negligible extragalactic magnetic field (0.1 and 1~nG). For these results, an iron only source composition with spectral index $\alpha=1.8$) and cutoff energy $E_{\rm Fe, max}=10^{21}$~eV, have been used.}
\label{Mag_Effects}
\end{center}
\end{figure}

Using the results shown in Fig.~\ref{Si_Fe_shells_Mag}, the effect introduced into the results of 
section~\ref{First_Source} by the presence of a non-negligible ($>$pG) extragalactic 
magnetic field may be addressed. The dominant effect of these magnetic fields in combination 
with an absence of local sources is their alteration of the arriving composition, as demonstrated 
explicitly for $0.1$~nG and $1$~nG extragalactic magnetic fields in Fig.~\ref{Mag_Effects}. Indeed,
with a better handle on both the composition and distance to the nearest source, the composition
may also provide a valuable probe of the local (intervening) extragalactic magnetic field.

\section{Summary}
\label{Summary}

% Previous work on this topic
Following the motivation that UHE CRs consist of heavy nuclei, whose
sources have been suggested to be local~\cite{Wibig:2007pf,Piran:2010yg}, we 
have here looked closer at the requirements on their source distribution.
In this work we obtained concrete quantitative constraints on the
UHE CR source population. Making explicit use of the recently provided Auger 
spectral and shower composition 
results, along with detailed UHE CR nuclei modeling, we investigated whether consistency
may be found with this data using single source composition models. 

For the case of negligible extragalactic magnetic fields, we have demonstrated that
a simplified analytic description agrees well with the spectrum and composition 
results obtained from the complete Monte Carlo description. 
Utilising this analytic description, we made a scan over the source spectral index and 
exponential energy cutoff, for a single source composition scenario, to obtain the goodness-of-fit 
contours. Taking into account the systematic errors in the flux and composition measurements, we found that hard ($\alpha<2$) source spectral indices and intermediate
cutoff energies ($E_{\rm Fe, max}\sim 10^{20.5}-10^{21}$~eV) for intermediate-to-heavy nuclei 
could provide a good fit to the full set of Auger UHE CR measurements above 10$^{19}$~eV.

Through the consideration of shells of UHE CR sources, we investigated the proximity of 
the UHE CR nuclei sources required if the presently observed trend of an increasingly 
heavy composition continues up to the highest energies observed by the ground array 
($10^{20.2}$~eV). By varying the size of the local void up to the nearest source, 
we investigated how detrimental the effect of this was on the goodness-of-fits
contours. Provided that the sources maximum energy lay below $10^{22}$~eV, we found
that the nearest sources had to be within $60$~Mpc and $80$~Mpc for silicon and
iron only sources respectively. 

If, however, extragalactic magnetic fields are sufficiently strong ($>$pG), the 
arriving flux from the different nuclei source shells is altered considerably. Though 
this effect is weakest at the highest energies, a local void of sources scenario 
with such an intervening field, may alter the arriving flux at energies below the 
cutoff feature, and thus alter the goodness-of-fit contour landscapes obtained in
Fig.~\ref{Contours}. However such ($<$nG strength) fields are found to be unable
to alter significantly our upper bound on the nearest source distance.

The requirement for local candidate objects able to satisfy the Hillas criterion \cite{Hillas:1985is} 
for $>10^{20}$~eV nuclei, whilst at the same time not disintegrating these nuclei during the
acceleration process, places a non-trivial dual condition on the source
environment. 
%Added to this the requirement that an overabundance (relative
%to that of the nearby ISM) of heavy nuclear species 
%are injected into the accelerator further limits the range of possible 
%candidates. 
At first consideration both the required proximity of the sources and their
required nuclei tolerant radiation fields would seem to exclude GRBs as viable candidates.
However, recent motivations for a Galactic GRB-type origin \cite{Calvez:2010uh} 
and the possibility that heavy nuclear species are produced during the GRB 
explosion \cite{Metzger:2011xs} demonstrate that the case for a GRB origin 
remains viable.
Due to the dependence of photo-disintegration rates on the radiation field level, 
acceleration sites far away from the central engine are naturally favoured. 
In this regards, acceleration within AGN jets or their radio lobes \cite{O'Sullivan:2009sc} 
are able to satisfy the above mentioned dual condition. 

With regards energetics, the UHE CR source luminosity density required to power the 
UHE CR population above $10^{19}$~eV is $\sim 5\times 10^{44}$~erg~Mpc$^{-3}$~yr$^{-1}$ 
\cite{Waxman:1995dg}. For the hard sources spectra motivated in this paper, approximately
the same source luminosity density would exist up to the cutoff energy. Thus an average 
local luminosity per source of 
$\sim 10^{43}/N_{\rm s}$~erg~s$^{-1}$ is required within the local 60~Mpc region, where 
$N_{\rm s}$ is the number of contributing sources. Such a luminosity is roughly $1/N_{\rm s}$ 
that of the 20-40~keV X-ray luminosity of local AGN \cite{Beckmann:2005gs,Baumgartner:2010}, 
of which $\sim$10 sit within 60~Mpc from Earth.
Whether it is reasonable for these local AGN to channel such a large fraction of their 
non-thermal luminosity into UHE CR flux, and whether a sufficiently enhanced nuclear 
composition is able to be achieved within their acceleration sites, however, remains 
unclear \cite{Pe'er:2009rc,GopalKrishna:2010wp}.

Our results demonstrate that exciting consequences follow from the intermediate-to-heavy 
nuclei component uncovered by Auger measurements. With nuclei photo-disintegration 
inevitably occurring during propagation, tough constraints are placed on the source 
proximity and environment. 
Furthermore, future Auger spectral and composition measurements are anticipated
to soon tighten these constraints.
With few candidate sources within the present upper bound distance suggested ($<60$-$80$~Mpc), 
the puzzle as to the UHE CR origin both remains and becomes even more intriguing.

\acknowledgements{AMT acknowledges support by the Swiss National Science Foundation grant PP00P2 123426. MA acknowledges support by the US National Science Foundation Grant No PHY-0969739 and by the Research Foundation of SUNY at Stony Brook.
}

\begin{appendix}

\section{Goodness-of-Fit Test}\label{gof_test}

We perform a goodness of fit (GOF) test of the compatibility of the Auger data with a given model following Ref.~\cite{Ahlers:2010fw}. For a fixed source composition of nuclei we vary the universal spectral index $\alpha$ of the emission spectrum and the maximal rigidity cutoff $E_{\rm Fe, max}$, described in \footnotemark[\value{pub}] and \footnotemark[\value{pub2}].

Given the acceptance $A_n$ (in units of area per unit time per unit solid angle) of the experiment in bin $n$ centered at energy $E_n$ with bin width $\Delta_n$, the number of expected events is
\begin{multline}
\mu_n(\alpha,E_{\rm Fe, max},{\cal N},\delta_E)\\= A_n{\cal N}
\int\limits_{E^-_n}^{E^+_n}
{\rm d}E\, \sum_i\frac{{\rm d}N_i}{{\rm d}E}(E)\,,
\label{eq:nev}
\end{multline}
where ${\rm d}N_i/{\rm d}E$ is the flux of nuclei $Z$ arriving at the detector and the boundaries are $E^\pm_n = E_n(1+\delta_E)\pm\Delta_n/2$. The parameter $\delta_E$ in Eq.~\eqref{eq:nev} is a fractional energy-scale shift that takes into account the uncertainty in the energy-scale and ${\cal N}$ is the normalization of the source luminosity.

A fraction $\epsilon_n$ of the expected $\mu_n$ events per bin passes the quality cuts for the distribution of shower maxima $X$. This distribution is only known by its first two moments, the average shower maximum $\langle X\rangle_n$ and its root-mean-square (RMS) $\Delta X_n$~\cite{Abraham:2010yv}. For the GOF we hence divide the distribution of shower maxima $X$ into three intervals $(X^-_{m,n},X^+_{m,n})$, such that each interval contains $1/3$ of the total number of observed events. For a Gaussian distribution this corresponds to a central bin of size $X^\pm_{0,n} \simeq \langle X\rangle_n(1+\delta_{\langle X\rangle})\pm0.43\Delta X_n(1+\delta_{\rm RMS}))$ and two bins containing all other events left ($m=-1$) and right ($m=1$) to it. Here, we introduce the fractional uncertainties $\delta_{\langle X\rangle}$ and $\delta_{\rm RMS}$ of the mean and RMS of the shower maximum, respectively. 

\begin{figure}[t]\centering
\includegraphics[width=0.9\linewidth,type=pdf,ext=.pdf,read=.pdf]{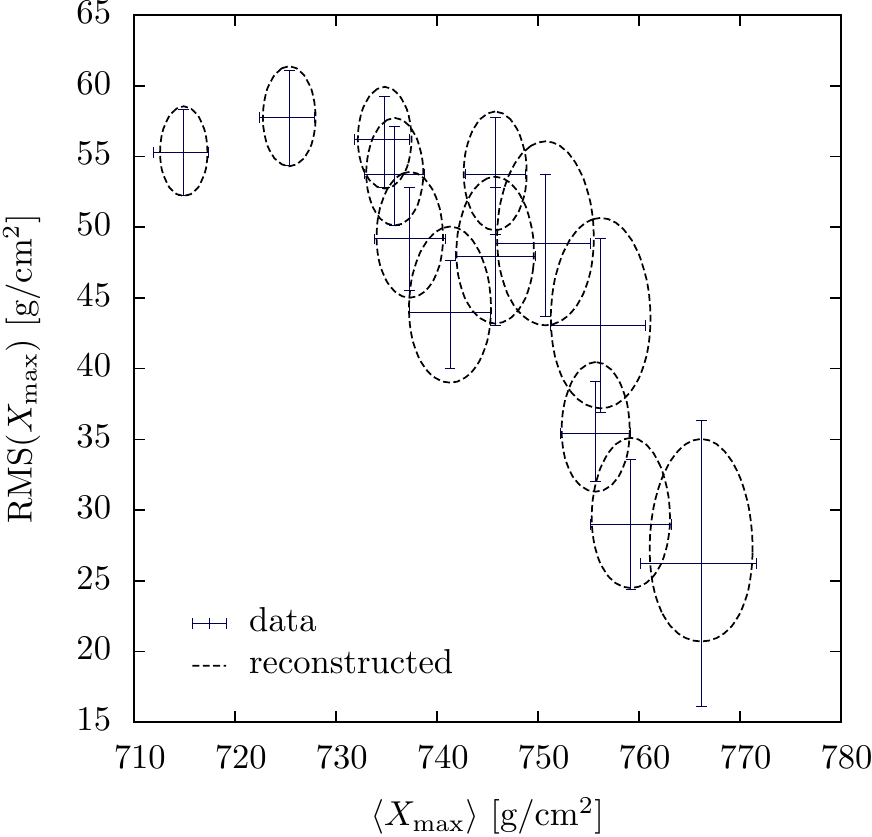}
\caption[]{The statistical uncertainty of the measured mean and RMS of the shower maximum from Ref.~\cite{Abraham:2010yv} compared to the uncertainty of the reconstruction from the three $X$-bins introduced for the goodness of fit.}\label{reconstruction}
\end{figure}

We emphasise that the binning method of the shower maxima that we employ can only as an approximation. 
As a check of this approach we try to reconstruct $\langle X\rangle$ and $\Delta X$ from this X-binning by a simple $\chi^2$-fit. The number of events passing the quality cuts is known from Fig.~2 of Ref.~\cite{Abraham:2010yv}. Figure~\ref{reconstruction} shows the reconstructed 1$\sigma$ contours in comparison to the statistical uncertainty of the measurement extracted from Fig.~2 and 3 of Ref.~\cite{Abraham:2010yv}. The contours are consistent with the data and reproduce the right order of magnitude of the statistical uncertainty.

The predicted distribution of shower maxima can be determined by hadronic interactions models such as QGSJET~11 \cite{qgsjet_11}. We assume in the following that the distribution in the $n$-th energy bin can be approximated by a Gaussian distribution with mean $X_n$ and RMS $\sigma_n$. The expected number of events in the X sub-bins $m=-1,0,1$ is hence of the form
\begin{multline}
\mu_{m,n}(\alpha,E_{\rm Fe, max},{\cal N},\delta_E,\delta_{\langle X\rangle},\delta_{\rm RMS})\\= \epsilon_n \mu_n(\alpha,E_{\rm Fe, max},{\cal N},\delta_E)\int\limits_{X^-_{m,n}}^{X^+_{m,n}}
\!\!{\rm d}X\, \frac{e^{-\frac{(X-X_n)^2}{2\sigma_n^2}}}{\sqrt{2\pi}\sigma_n}\,.
\label{eq:nevx}
\end{multline}

The probability $P_n(N_{{\rm tot},n})$ of observing $N_{{\rm tot}, n}$ events in the $n$-th energy bin follows a
Poisson distribution $f(N_{{\rm tot}, n},\mu_n)$ with mean $\mu_n$. The $\epsilon_n N_{{\rm tot}, n}$ events passing the quality cuts distribute between the three X-bins following a Poisson distribution with mean $\mu_{m,n}$ subject to the constraint $\epsilon_n N_{{\rm tot},n} = N_{-1,n} + N_{0,n} + N_{1,n}$. The conditional probability is
\begin{multline}
P'_n(N_{-1,n},N_{0,n},N_{1,n})\\=\frac{f(N_{-1,n},\mu_{-1,n})f(N_{0,n},\mu_{0,n})f(N_{1,n},\mu_{1,n})}{f(\epsilon_nN_n,\epsilon_n\mu_n)}\,.
\end{multline}
Hence the total probability of observing a set of events $\lbrace N \rbrace$ is given as 
\begin{multline}
P({\lbrace N \rbrace};\alpha,E_{\rm Fe, max},{\cal N},\delta_{\langle X\rangle},\delta_{\rm RMS},\delta_{E})\\=\prod_n P_n(N_{{\rm tot},n})\prod_\ell P'_\ell(N_{-1,\ell},N_{0,\ell},N_{1,\ell})\,.
\end{multline}
Here the first product runs over all energy bins used for the CR spectrum and the second product runs over all bins where additional information about the elongation rate distribution is available.
 
As a next step we marginalize over the experimental systematic uncertainty in the energy scale $\delta_E$, shower maximum $\delta_{\langle X\rangle}$ and root-mean-square $\delta_{\rm RMS}$ as well as the normalization $\mathcal{N}$. We assume a flat prior with $|\delta_E|< 20$\%, $|\delta_{\langle X\rangle}|< 2$\% and $|\delta_{\rm RMS}|< 10$\%. Marginalization is done by maximizing the probability of the actual experimental observation. We hence define
\begin{multline}
\widehat P({\lbrace N \rbrace};\alpha,E_{\rm Fe, max})\\\equiv
P({\lbrace N \rbrace};\alpha,E_{\rm Fe, max},\widehat{\cal N},\widehat\delta_{\langle X\rangle},\widehat\delta_{\rm RMS},\widehat\delta_{E})\,,
\label{eq:margi}
\end{multline}
where the set of parameters $\lbrace\widehat{\cal N},\widehat\delta_{\langle X\rangle},\widehat\delta_{\rm RMS},\widehat\delta_{E}\rbrace$ maximizes the probability (\ref{eq:margi}) for the experimental result ${\lbrace N^{\rm exp} \rbrace}$, consisting of the total events per bin $N^{\rm exp}_{{\rm tot},n}$ as well as $N^{\rm exp}_{m,n} = \epsilon_nN^{\rm exp}_{{\rm tot}, n}/3$, if available, which follows from the construction of the X-binning.

The transition between the galactic and extragalactic component of UHE CRs is uncertain. A natural candidate is the CR ``ankle'' that is observed in the Auger data at about $10^{18.6}$~eV. We assume that the ``contamination'' by a low-energy galactic component has become negligible beyond $10^{19}$~eV and include only Auger data above this threshold for our GOF. For consistency we have to check that the extragalactic component does not over-shoot the CR data at lower energies. Hence, in the marginalization procedure we also impose a prior on the normalization ${\cal N}$ by requiring that the model spectra do not exceed the Auger data below the first bin used in the fit by more than three standard deviations. 

Finally, the model $(\alpha,E_{\rm Fe, max})$ is compatible with the experimental results at a given GOF if
\begin{equation}
\sum\limits_{\widehat P(\lbrace N \rbrace)>\widehat P(\lbrace N^{\rm exp} \rbrace)} 
\!\!\!\widehat P({\lbrace N \rbrace};\alpha,E_{\rm Fe, max})
\leq {\rm GOF}\,.
\end{equation}
Technically, this calculation is performed by generating a large number $\lbrace N^{\rm rep}\rbrace$ of replica experiments following the probability distribution (\ref{eq:margi}) and by imposing that a fraction $F$ with $\widehat P({\lbrace N^{\rm rep} \rbrace})>\widehat P({\lbrace N^{\rm exp} \rbrace})$ satisfies $F\leq{\rm GOF}$.

%%%%%%%%%%%%%%%%%
%%%%%%%%%%%%%%%%%

\section{A Simplified Description for UHE CR Propagation Through Turbulent Fields}
\label{turb_fields}

We here describe the results of an investigation we carried out comparing
two different descriptions for the diffusive
propagation process of UHE CR in turbulent magnetic fields. 

As discussed in \cite{Giacalone:1994}, a turbulent field may be generated 
by adding plane waves whose amplitudes are dictated by the magnetic power 
spectrum,
\begin{eqnarray}
\bfm{B}=\sum_{n}\delta \bfm{B}_{n}
\end{eqnarray}
where $\delta \bfm{B}_{n}=\bfm{\xi_{n}} A_{n}e^{\left(ik_{n}x'+\phi_{n}\right)}$, 
with $x'=x\cos\alpha-y\sin\alpha$, $\alpha$ describing the orientation 
of the wave (randomly chosen), $\bfm{\xi_{n}}$ describing the (randomly chosen)
polarisation of the wave (ie. over each wavelength ${\bfm{\xi_{n}}}$ describes an 
ellipse in the $y'-z'$ plane), 
$A_{n}$ describing the amplitude of the wave, $k_{n}$ ($=2\pi/\lambda_{n}$)
describing the wavenumber, and $\phi_{n}$ is a random phase.
The amplitudes of the different waves are given a power law distribution
of the form $A_{n}\propto k_{n}^{-q/2}$. For the purpose of our comparison
we assume a Kolmogorov spectrum ({\it i.e.}~$q=5/3$).

We prepare a turbulent magnetic field region, containing an ensemble of
waves with wavelengths between $\lambda_{\rm min}$ and $\lambda_{\rm max}$.
It is assumed that a guiding mean-field, $B_{0}$, lies along the 
z-direction, and that the sum of the energy
within the turbulent field is equal to the total energy within the guiding
field ($B_{0}^{2}=\sum_{n}\delta B_{n}^{2}$). We describe the turbulence
using $50$ isotropic waves, with $10$ waves per decade 
({\it i.e.}~$\lambda_{\rm max}/\lambda_{\rm min}=10^{5}$). 
The Bulirsch-Stoer method was used to track the particles in the 
field.

A simplified description of turbulent propagation, however, may also be
obtained through a scattering description of UHE CR interaction with the
turbulent field. In this description, UHE CR stochastically interact with 
scattering centers whose density is dictated by the energy density in the
turbulence power spectrum at the wavelength matching the gyro-radius of the
UHE CR. We refer to this description as the ``delta-approximation'' case,
since it assumes only resonant scattering. 
The resonant scattering angle of this process is given by
\begin{equation}
\Delta \theta = \begin{cases}1&R_{\rm L}<L_{\rm coh}\,,\\\frac{L_{\rm coh}}{R_{\rm L}}&R_{\rm L}>L_{\rm coh}\,, \end{cases}
\end{equation}
where $R_{\rm L}$ is the particle's Larmor radius and $L_{\rm coh}$ is the magnetic
field's coherence length. The corresponding scattering length is then given by
\begin{equation}
R_{\rm scatt}(\Delta \theta)=aL_{\rm coh}\begin{cases}\left(\frac{R_{\rm L}}{L_{\rm coh}}\right)^{2-q}&R_{\rm L}<L_{\rm coh}\,,\\1&R_{\rm L}>L_{\rm coh}\,,\end{cases}
\end{equation}
where the factor $a$ describes how far this scattering is from the Bohm
regime ({\it i.e.}~$a=1$ at the Bohm limit, giving rise to approximately one scattering 
per gyro-radius) for particles whose gyro-radii are at the scale $\lambda_{\rm max}$. 
The factor $(2-q)$ may be thought 
to describe the exponent in the energy dependence of the density of (resonant turbulence) 
scatterers. 
In order to scale down this result to allow us to probe the low angle
scattering regime, we employ a small angle scattering length of size $\delta\theta$, 
$R_{\rm scatt}(\delta\theta)=(\delta\theta/\Delta\theta)^{2}R_{\rm scatt}(\Delta\theta)$. 
Note, the intrinsic scattering in angle introduced through each photodisintegration process
is always heavily sub-dominant, and can be safely ignored.

\begin{figure}[ht]
{\includegraphics[angle=0,width=0.9\linewidth,type=pdf,ext=.pdf,read=.pdf]{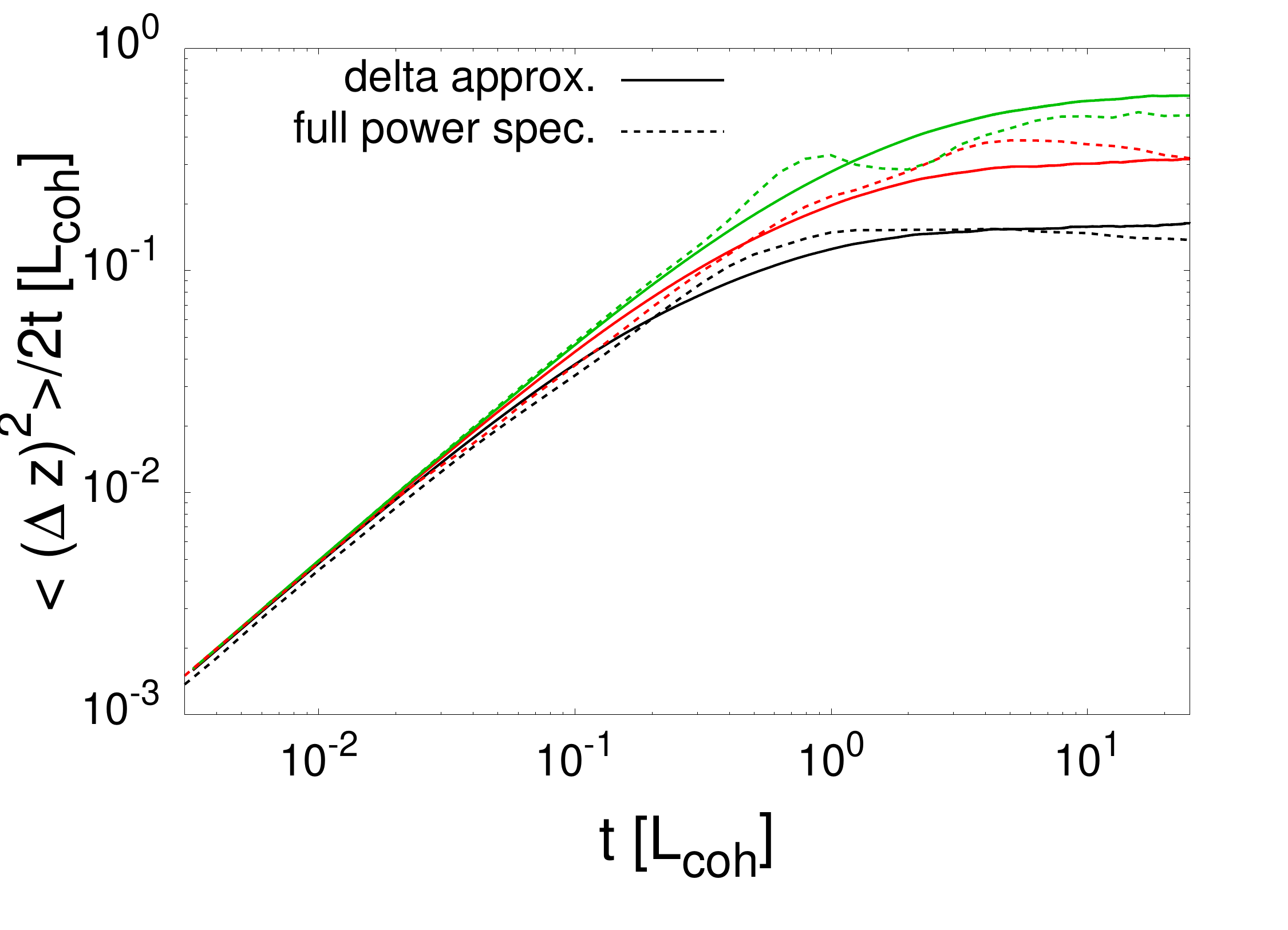}}
\caption{A plot comparing the spreading of particles along the mean-field direction during propagation for both the full turbulent field description and the simplified ``delta-approximation''. These results were obtained for 10$^{18}$~eV energy UHE CR iron nuclei in 0.1~nG, 1~nG, and 10~nG extragalactic magnetic field with a 1~Mpc coherence length.}
\label{diffusion_comparison}
\end{figure}

We compare the growth of spread of particles in the mean-field direction ($\Delta z$) as a
function of time for both the simplified description and that of the full turbulent 
field description in Fig.~\ref{diffusion_comparison}, for the case of $a=1$. For this simulation,
the particles were all started from the same position in the static turbulent field with an
isotropic distribution of directions.

With good agreement found between the ``full power spectrum'' and the ``delta-approximation''
descriptions we conclude that the ``delta-approximation'' method is able to provide an
accurate description for UHE CR propagation in extragalactic magnetic fields.

\end{appendix}

\end{document}